# The hybrid lattice of $K_xFe_{2-y}Se_2$: why superconductivity and magnetism can coexist

Despina Louca[1*], Keeseong Park[1#], Bing Li[1], Joerg Neuefeind[2], JiaqiangYan[2,3]

[1] *Department of Physics, University of Virginia, Charlottesville, VA 22904, USA*

[2]*Oak Ridge National Laboratory, Oak Ridge, TN 37831, USA*

[3]*Department of Materials Science and Engineering, The University of Tennessee, Knoxville, TN 37996, USA*

**Much remains unknown of the microscopic origin of superconductivity when it materializes in atomically disordered systems as in amorphous alloys[1] or in crystals riddled with defects[2]. A manifestation of this conundrum is envisaged in the highly defective iron chalcogenide superconductors of $K_xFe_{2-y}Se_2$[3-6]. How can superconductivity survive under such crude conditions that call for strong electron localization[7]? With vacancies present both at the K and Fe sites, superconductivity is bordering a semi-metallic region below x ~ 0.7 and an insulating and antiferromagnetic region above x ~ 0.85[8,9]. Here, we report on the bulk local atomic structure and show that the Fe sublattice is locally distorted in a way that it accommodates two kinds of Fe valence environments giving rise to a bimodal bond distribution. While the bond length distribution is driven by K and Fe contents, the superconducting state is characterized by the coexistence of both short (metallic) and long (insulating) Fe bond environments and is not phase separated. In contrast to other Fe-based materials in which only one kind of Fe to Fe bond is present, the dual nature of the**

**Fe correlations explains why superconductivity is intertwined with magnetic order. Such a hybrid state is most likely present in cuprate superconductors as well[10,11] while our results point to the importance of the local atomic symmetry by which the exchange interactions between local moments can materialize[12].**

While suppression of long-range magnetic ordering has long been thought to be a pre-requisite to the onset of superconductivity, this rule is broken in the new class of iron-based superconductors, $K_xFe_{2-y}Se_2$, with $T_c$'s of the order of 30 K or so[8,9]. Sandwiched between bad metallic ($x$ < ~0.7) and insulating ($x$ > ~0.8) antiferromagnetic states, the superconducting phase exists in tandem with large magnetic moments with a high antiferromagnetic transition, $T_N$, close to 560 K[13]. Even more peculiar is the size of the magnetic moment which is over $3\mu_B$ per Fe atom, a non-trivial quantity, and much larger than what is found in other pnictides and chalcogenides. Distinct to this system is the proposed presence of iron vacancies at one of the two possible crystal sites and of their ordering that gives rise to the $\sqrt{5}\times\sqrt{5}\times1$ type ordered structure with the *I4/m* crystal symmetry[8, 14-16]. A vacancy order-disorder transition occurs at $T_S$ ~ 500 K[5], accompanied by a change in the crystal symmetry to *I4/mmm*. Are the vacancies ordered or disordered or even present in the superconducting state? It is currently argued that of the systems that exhibit a superconducting transition, phase separation occurs between the superconducting and non-superconducting regions of the sample, and that the superconducting component is stoichiometric with the $KFe_2Se_2$ composition and has no Fe vacancies[13], while the insulating component has the $\sqrt{5}\times\sqrt{5}\times1$ vacancy ordered structure. The Fe vacancies are considered destructive to superconductivity[17]. What is the interplay between magnetism, superconductivity and vacancies and do they microscopically coexist or are they phase separated? Here we show by probing the local atomic structure via neutron diffraction that the defect structure is not phase separated. Instead, a hybrid state is most likely present that accommodates two types of Fe bonds that are tuned continuously by changing the composition and temperature.

To identify the local symmetry that governs the Fe lattice, two different crystal structure scenarios are considered. The first consists of the *I4/m* crystal symmetry shown in Fig. 1(a) where two distinct iron sites are present labeled as Fe1 and Fe2, and give rise to three unique bond lengths due to Fe1-Fe1, Fe1-Fe2, and Fe2-Fe2 pairs. Details of the structure coordinates are provided in a supplemental table. The Fe1 site is the one that is partially occupied. In the second scenario, the *I4/mmm* crystal symmetry is considered which is the symmetry observed in the pnictides of $Ba(Fe_{1-x}Co_x)_2As_2$ where only one unique Fe is present[18,19]. This symmetry does not hold any Fe vacancies and may be the symmetry appropriate for the superconducting phase if a phase-separated mixture existed. These two scenarios can be directly tested by considering the local Fe environment under the constraints of the two symmetries. Determined via the pair density function (PDF), $\rho(r)$, analysis technique, the local Fe to Fe correlations alone are shown in Fig. 1(c) with the x-axis plotted in real-space. Given that the *I4/mmm* symmetry accommodates only one Fe site, only one Fe-Fe bond is seen at distances between 2.6 - 3 Å. Shown in descending order are the partial functions for the Fe2-Fe2, Fe1-Fe2 and Fe1-Fe1 correlations calculated using the *I4/m* symmetry. Since this symmetry holds two Fe sites, the combined total partial function for Fe has two peaks at distances between 2.6 - 3 Å. For comparison, shown in Fig. 1(d) are plots of the square Fe-Fe plaquettes in $FeSe_{1-x}Te_x$[20] identifying the singular Fe-Fe bond length at 2.69 Å (in the *P4/nmm* symmetry) and at 2.81 Å in $Ba(Fe_{1-x}Co_x)_2As_2$[18,21] (in the *I4/mmm* symmetry).

The evolution of the crystal structure in $K_xFe_{2-y}Se_2$ is determined by investigating chemical compositions in the three distinct regions of the phase diagram: (I- high Fe valence) x = 0.6, (II – intermediate Fe valence) x = 0.8, and (III – low Fe valence) x = 1.0[13], only one of which is superconducting, the $x$ = 0.8, with a $T_C$ of about 30 K. Its bulk susceptibility is shown in Fig. 1(b). The experimentally obtained $\rho(r)$ function is determined by Fourier transforming the total structure function to high momentum transfers (42 $Å^{-1}$). Shown in Fig. 2(a) is the data corresponding to the local structures for the three compositions at 2 K while in 2(b), the data for the same compositions were collected at 200 K. With the addition of potassium K ions, the function shifts to the right because the lattice expands with the addition of more K. Apart from this however, differences among the three functions are observed across real space. Most importantly, in the region where the Fe-Fe bonds can be clearly viewed as they are isolated from the rest of the structure, two peaks are present (indicated by arrows), albeit at different concentrations, that closely resemble what is expected from the *I4/m* symmetry (see Fig. 1(c)). As $x$ increases from 0.6 to 1.0, the distribution of short and long Fe-Fe bond lengths continuously shift from the short at ~ 2.68 Å to the long at ~ 2.87 Å. It is interesting to compare to other Fe systems such as metallic complexes where metallic Fe-Fe bonds are of the order of 2.64 Å[22] while in metallic ferromagnetic Fe, the Fe-Fe bond is 2.48 Å long. On the other hand in the antiferromagnetic insulating $\alpha$-$Fe_2O_3$, with the corundum structure, the Fe-Fe bonds are 2.95 Å in length. In the present case, most clearly is the presence of both short and long Fe-Fe bonds in the superconducting composition, right in-between the other two compositions. At 200 K in Fig.

2(b), a redistribution of the weight under the short Fe-Fe peak is observed as it shifts to the longer length with increasing $x$ (Fig. 2(b)). Focusing on the $x = 0.8$ superconductor as a function of temperature, the width of the first Fe-Fe peak increases as its height decreases (see inset, red symbols), while the second Fe-Fe peak gains a bit in height (see inset, blue symbols) as it slightly shifts to the left. Back to the composition dependence, at 2 K, the region under the two Fe-Fe peaks is fit by two Gaussians and integrated to obtain the areas shown in Figs. 2(d)-(f). The continuous shift from the short to the long bonds as a function of $x$ is evident here as well, but at all times, both peaks are present. This serves as evidence for a continuous structural evolution with doping.

How can we describe the local microstructure of the superconducting state? The current high resolution PDF study reveals local bond-length inhomogeneities which are not apparent in the average crystal structure. The double-well bond length distribution can be explained well by a linear superposition of the local structures obtained in regions (I) and (III). This is shown in Fig. 3(a). The valence state of Fe has a direct effect on the Fe-Fe bond length, with the bond length increasing with increasing potassium doping. The double Fe-Fe distribution indicates that two Fe valence environments are present. The shift to the longer length with $x$ indicates that the valence of Fe goes down while the shift to the shorter length indicates that the Fe valence goes up. It was previously noted that the Seebeck coefficient changes systematically with the Fe valence which corresponds to a change in the type of charge carriers[13]. The superconducting composition shows both Fe environments that may explain why its Seebeck coefficient is small as the negative and positive values cancel each other out, leading to a Lifshitz transition[23]. At the same time, atomically, the $x = 0.8$ superconductor is not a phase separated state as no double diffraction peaks were observed but rather evolves continuously by doping K into the vacancy ordered $\sqrt{5} \times \sqrt{5} \times 1$ structure. A comparison of the $x = 0.8$ PDF with a model calculated based on the *I4/mmm* symmetry clearly shows that there is no fully stoichiometric phase present in this system as previously suggested, as this will generate an intermediate Fe-Fe peak (Fig. 3(b) – blue line). A model comprised of a linear combination of the *I4/mmm* and *I4/m* structures does not compare well with the superconducting state either. Second, if we were to assume that there are two *I4/m* phases where in one the Fe1 site is fully occupied while in the other phase, the Fe1 site is partially empty, the combined model again cannot reproduce the local Fe-Fe bond distribution. This is because the short Fe correlations are enhanced by combining the two phases which is clearly not what is observed locally.

How well the average $\sqrt{5} \times \sqrt{5} \times 1$ structure fits the local symmetry of the superconductor can be seen in Fig. 3(b). Even though the $\sqrt{5} \times \sqrt{5} \times 1$ structure yields two Fe-Fe peaks, their intensity ratio does not correspond to the double-well Fe bond distribution observed experimentally. To create a bimodal Fe valence environment and reproduce the experimental configuration observed in the $x = 0.8$, it requires us to expand the inner square Fe2. It turns out that the Fe2-Fe2 bonds are the majority at ~2.68 Å while the longer Fe2-Fe2 bonds at~ 2.87 Å bonds are the minority in the $\sqrt{5} \times \sqrt{5} \times 1$ structure. The bonds between Fe1-Fe2 contribute little to the overall count because of the low concentration of Fe1. Thus to increase the length from short to long requires that the inner Fe square plaquette expands from the configuration shown in Fig. 3(c) which corresponds to the $x = 0.6$ structure to the configuration shown in Fig. 3(d) for the $x = 1.0$ structure. Such a local lattice mode will initially appear in some parts of the sample, but as the concentration of K increases or temperature decreases, it spreads to more sites.

What determines the critical concentration of $x = 0.8$ may well rest in this local arrangement that in turn leads to changes of the Fermi surface. Having two Fe environments may also explain the intertwining of magnetic and superconducting orders which may be more frequent in nature than previously anticipated, as recent reports of a charge-density wave order in the superconducting state of $YBa_2Cu_3O_{6+y}$ are revealing[24,25].

In other Fe-based superconductors, it has been shown that the Fermi surface consists of hole and electron pockets while nesting may lead to a spin density wave[26]. However, $K_xFe_{2-y}Se_2$ is devoid of hole pockets at the zone center while strong antiferromagnetic ordering is present instead, that cannot be explained by nesting[27-29]. From the structure point of view, this system is distinctly different from the $FeSe_xTe_{1-x}$ class with $T_C$'s ranging from 8-14 K, where only one distribution of Fe-Fe bond length is observed. Similarly, in the pnictide $Ba(Fe_{1-x}Co_x)_2As_2$ only a single Fe-Fe bond type is observed. While in both of these systems, no static long-range AFM order is present, this is in contrast to the K system, where all samples exhibit static antiferromagnetic order that can in turn be explained by the presence of the long Fe-Fe bonds. The presence of vacancies and local disorder lower the lattice symmetry that must in turn influence the electron-phonon interactions and $T_C$[30]. It is conceivable that such a disorder would weaken the coupling constant leading to a change in the electron and phonon density of states.

**Acknowledgements** The work at the University of Virginia and at Oak Ridge National Laboratory have been supported by the U. S. Department of Energy, Office of Basic Energy Sciences.

#Current address: Basic Science College, Daegu Gyeongbuk Institute of Science and Technology (DGIST), Daegu 711-873, Korea

**Author Information** Correspondence and requests for materials should be addressed to D. Louca (louca@virginia.edu)

## Methods

The $K_xFe_{2-y}Se_2$single crystals used in this study were grown by a modified Bridgman technique. The FeSe precursor, prepared separately, was mixed with pieces of K metal. The mixture was sealed in a glass tube in a glove box under Ar atmosphere and heated up to 1030°C for 12 hours. After 4 hours held at that temperature, the sample was cooled down to 730°C at 5K/min and then quenched in air by turning off the power to the furnace. Three nominal compositions were prepared: $K_{0.6}Fe_{2-y}Se_2$, $K_{0.8}Fe_{2-y}Se_2$ and $KFe_{2-y}Se_2$.The bulk magnetic susceptibility measurements from 5 to 50 K showed that $T_C$~ 30 K for the $K_{0.8}Fe_{2-y}Se_2$crystals only, while no superconductivity was observed for the other two compositions. The neutron experiments were carried out using the NOMAD diffractometer at the Spallation Neutron Source of Oak Ridge National Laboratory as a function of temperature, from 2 to 300 K. The single crystals were ground into powder inside a dry glove box prior to the experiment and held in an evacuated vanadium can in a displex refrigeration system.

## Figure captions

**Fig. 1 Structure and magnetism.** (a) The crystal structure of $K_xFe_{2-y}Se_2$ in the *I4/m* symmetry with vacancies. (b) The temperature dependence of the magnetization from zero-field-cooled (ZFC) and field-cooled (FC) measurements at a magnetic field of 10 Oe for the $x = 0.8$. (c) (upper) In the *I4/m* crystal structure, three kinds of Fe to Fe correlation functions are present based on the symmetry due to the two Fe positions. Also shown is the total Fe partial function in this symmetry. (lower) In the *I4/mmm* crystal structure, only one kind of Fe to Fe correlation function is present. The parameters for the two crystal symmetries are provided in a supplemental Table. (d) Projecting on the *ab*-plane, only one Fe-Fe bond length is present in $FeSe_{1-x}Te_x$ (upper) and in $Ba(Fe_{1-x}Co_x)_2As_2$ (lower).

**Fig. 2 Experimental local structure.** The pair density function is determined from the neutron data collected from the $x = 0.6$, 0.8 and 1.0 crystals at 2 K in (a) and 200 K in (b). Differences are observed across real space, especially in the region from 2.6 to 3.0 Å where two Fe to Fe correlations (long and short) are present. (c) The temperature dependence of the local structure corresponding to the superconducting x = 0.8. The width of the short Fe-Fe peak broadens with increasing temperature due to an increase in the short Fe-Fe bond distribution. In the inset, the peak height for the short (red) and long (blue) Fe-Fe bonds are plotted. The error bars are one standard deviation. (d)- (f) In the region from 2.6 to 3.0 Å of the ρ(r), the short and long Fe-Fe peaks are fit by two Gaussians functions. The ratio of the integrated intensity corresponding to the long and short Fe bonds is shown in each panel. Marked on (e) with an arrow is the possible position of the Fe-Fe peak had the *I4/mmm* symmetry been present.

**Fig. 3 Local structure environment.** (a) The experimentally determined pair density function for the $x = 0.8$ is compared to a linear superposition of the structures corresponding to the $x = 0.6$ and 1.0 at 2 K. (b) The pair density function for the $x = 0.8$ is compared to two models, calculated using the parameters of the *I4/m* and the *I4/mmm* crystal symmetries. (c) and (d) are projections on the *ab*-plane showing the Fe-Fe bonds for the $x = 0.6$ (c) and for the $x = 1.0$ (d) compositions. The superconducting phase is a hybrid of the two. From the $x = 0.6$ to the $x = 1.0$ structure, the inner Fe2 square lattice expands, reducing the number of short bonds while increasing the number of long Fe bonds.

**Supplemental information**

Table 1: Crystal structure information on the two compositions surrounding the superconducting phase.

| | **Model for $K_{0.6}Fe_{2-y}Se_2$ a = 8.5469 Å, c =14.1422 Å** | | | | **Model for $KFe_{2-y}Se_2$ a = 8.8389 Å, c = 13.8909 Å** | | | |
|---|---|---|---|---|---|---|---|---|
| | x | y | z | Occupation | x | y | z | Occupation |
| K1 | 0 | 0 | 0 | 1 | 0 | 0 | 0 | 1 |
| K2 | 0.404 | 0.178 | 0 | 0.6 | 0.395 | 0.197 | 0 | 0.98 |
| Fe1 | 0 | 0.5 | 0.25 | 0.06 | 0 | 0.5 | 0.25 | 0.4 |
| Fe2 | 0.203 | 0.0089 | 0.2508 | 1 | 0.21 | 0.089 | 0.2508 | 1 |
| Se1 | 0.5 | 0.5 | 0.1337 | 1 | 0.5 | 0.5 | 0.145 | 1 |
| Se2 | 0.1147 | 0.3011 | 0.1455 | 1 | 0.115 | 0.294 | 0.142 | 1 |

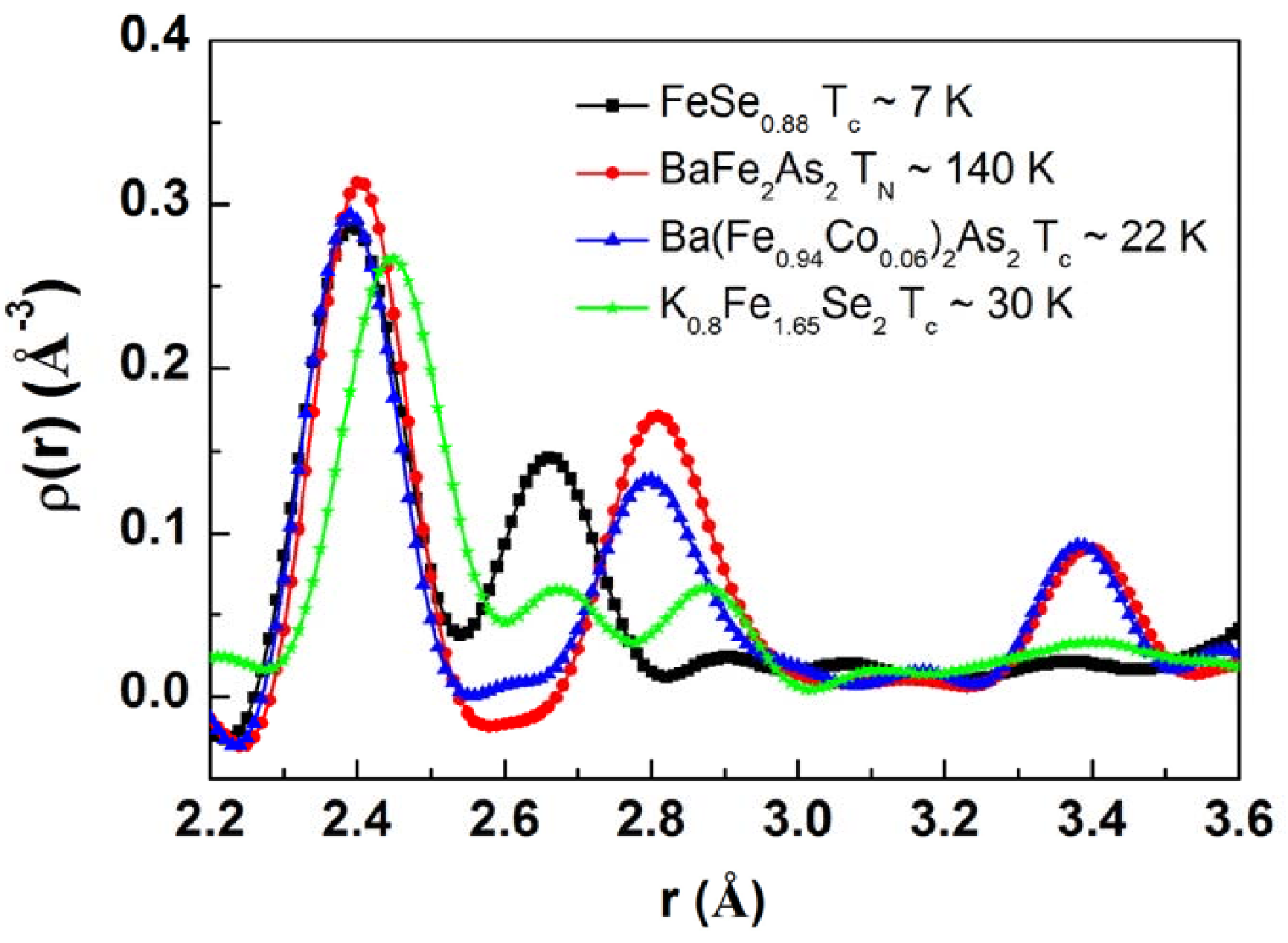

Supplemental figure, Louca et al.

Comparison of the local structures of 4 different compounds: FeSe, $BaFe_2As_2$, $BaFe_{1.88}Co_{0.12}As_2$ and $K_{0.8}Fe_{1.4}Se_2$. The first peak corresponds to Fe-Se/As correlations. The second corresponds to Fe-Fe correlations. The Fe correlations do not overlap with other pairs. Only in the $K_{0.8}Fe_{1.4}Se_2$ are two distinct Fe-Fe bonds observed that can foster two kinds of Fe correlations to develop. If comparing FeSe to $K_{0.8}Fe_{1.4}Se_2$, it can be seen that while the first Fe-Fe bond appears at about the same location for the two samples, a very distinct long Fe-Fe bond is present in $K_{0.8}Fe_{1.4}Se_2$. Similarly, a comparison between the Ba-122 samples and $K_{0.8}Fe_{1.4}Se_2$ shows that the Ba-122 have one very broad Fe-Fe bond distribution consisting of long and short while in $K_{0.8}Fe_{1.4}Se_2$, the distribution clearly splits.

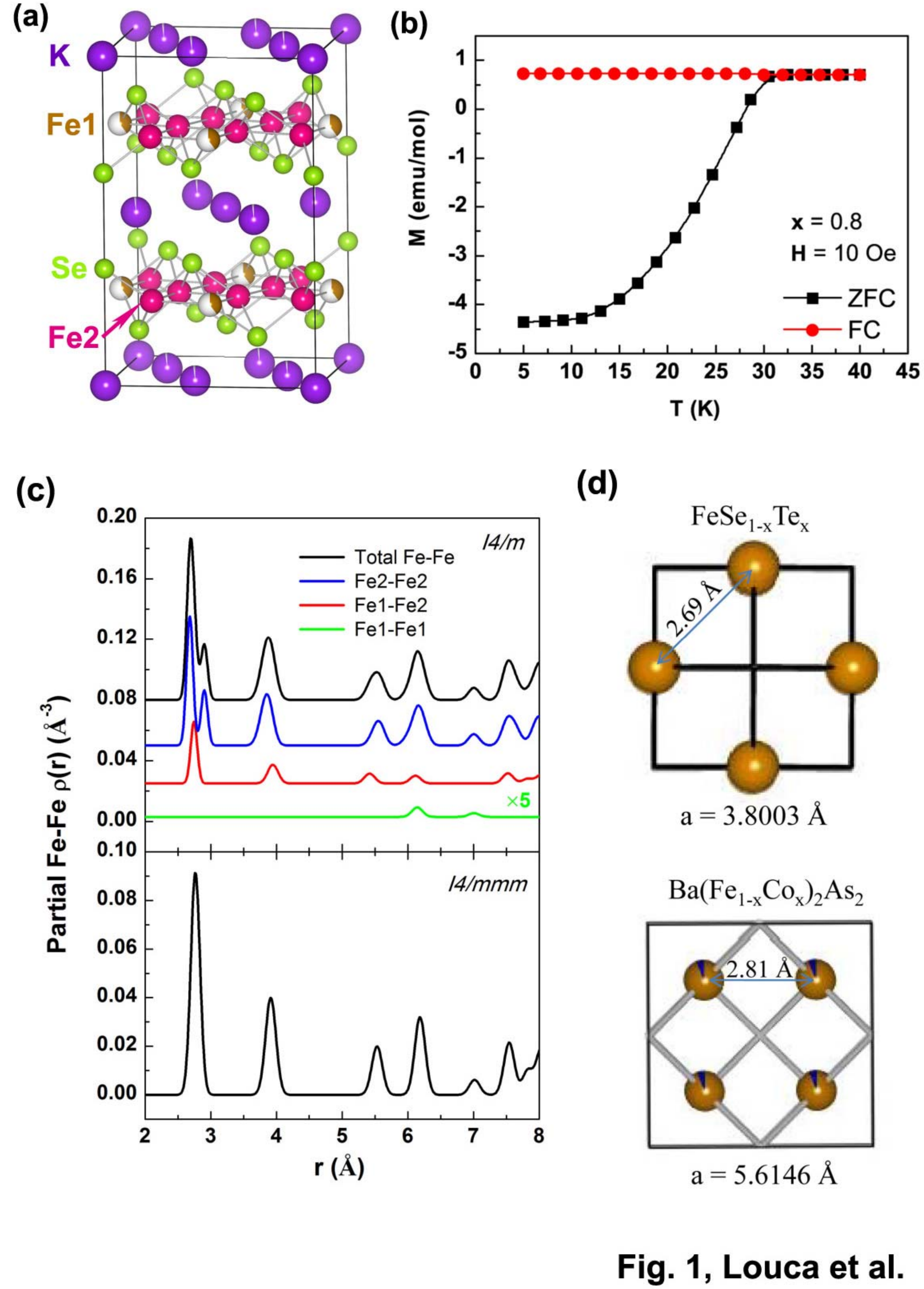

**Fig. 1, Louca et al.**

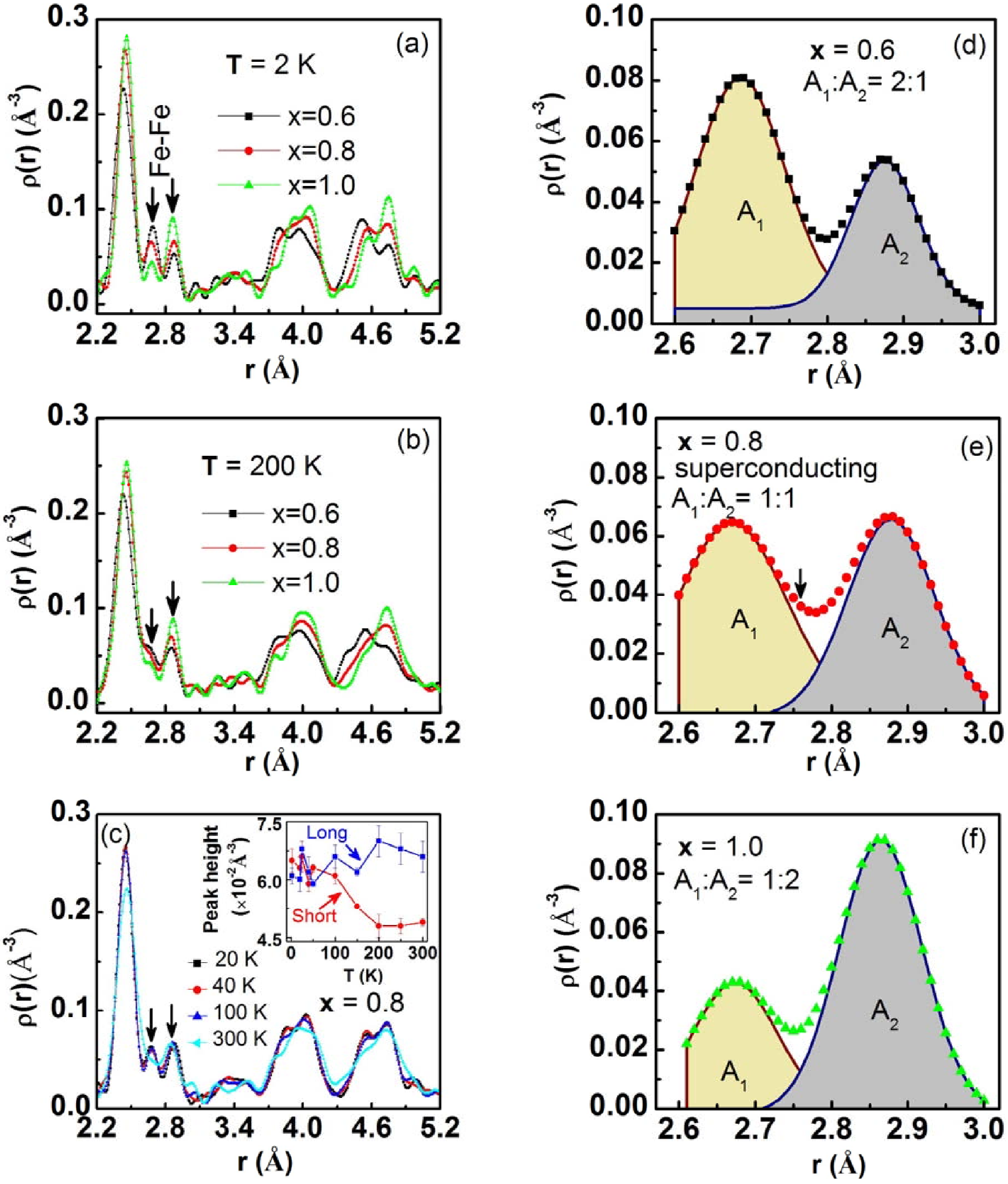

**Fig. 2, Louca et al.**

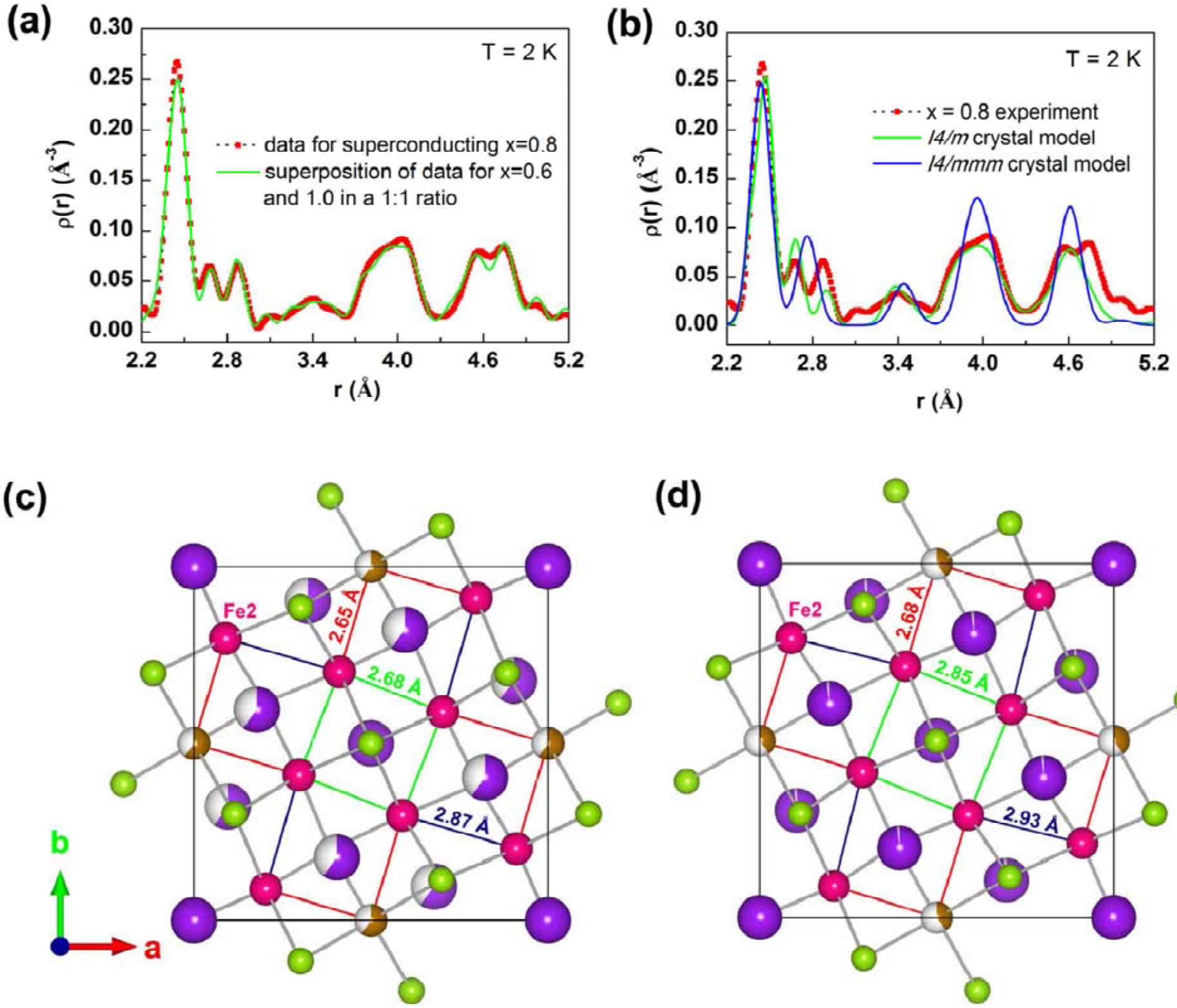

**Fig. 3, Louca et al.**